# Numerical Study of Kα X-ray Emission from Multi-layered Cold and Compressed Targets Irradiated by Ultrashort Laser Pulses

Hamed Koochaki Kelardeh
Department of Physics and Astronomy
Georgia State University, Atlanta, GA
hkoochakikelardeh1@physics.gsu.edu

Mohammad Mahdieh
Department of Physics
Iran University of Science and Technology (IUST)
mahdm @ iust.ac.ir

*Abstract*— In this paper the generation of Kα X-ray produced by interaction of ultrashort laser pulses with metal targets has been studied numerically. Several targets were assumed to be irradiated by high intensity ultra-short laser pulses for the calculations. Using Maxwell Boltzmann distribution function for hot electron and applying an analytical model, the number of Kα photons were calculated as a function of hot electron temperature, target thickness and K-shell ionization cross section. Also, simulation results of Kα yield versus target thickness variations from two and three layer metals have been presented. These calculations are useful for optimization of X-ray yield produced by irradiation of metal targets with high intensity laser pulses. We also generalized this model and present simulation results on Kα fluorescence measurement produced by fast electron propagation in shock compressed materials.

## I. Introduction

Recent developments of the chirped pulse amplification technique have given access to a new regime of laser-matter interactions with very intense laser fields [1-4]. The focusing of an ultra-intense laser beam on a solid target produces plasma on its surface [5-7]. Hot electrons are generated via collective absorption mechanisms, such as resonant absorption (RA) [8] or vacuum heating (VH) [9]. While the less energetic electrons deposit their energy in a thin front layer resulting in strong heating [10], the more energetic electrons penetrate much further into the target up to the colder regions behind the hot plasma where they ionize the k shell of the atoms giving rise to the emission of "cold" Kα photons. Since the electrons are generated only during the interaction with the laser pulse, a very short Kα pulse of the order of the laser pulse duration is expected [11-13]. The short duration makes these x-ray pulses very attractive for probing of matter dynamics on the femtosecond scale [14-16]. Moreover, because of its small x-ray emission size, it has a number of interesting applications for medical imaging techniques [17,18]. Further developments of these new x-ray sources are still needed before they can be used in practical applications.

The control and optimization of the x-ray emission of plasmas by high intensity laser-solid interaction is a subject of current interest. This requires an understanding of several mechanisms: the laser energy absorption, the hot electron generation, and the x-ray conversion. Several groups have already reported x-ray emission experiments relying on sub-picosecond laser systems. For example, Yu *et al.* systematically studied the hard x-ray emission produced by 500 fs laser pulses and obtained the intensity scale laws [19]. Soom *et al.* reported high Si and Al Kα conversion yield with 1.3 ps laser [20]. Eder *et al.* reported the observation of a maximum in Kα x-ray emission when the target is placed away from best focus [21]. Zhidkov *et al.* studied pre-pulse effect with a 42 fs laser [22] at moderate intensities and low contrast. This work demonstrated a decrease of the laser energy absorption for this kind of ultrashort pulse durations and also reported the critical influence of the plasma gradient for hot electron generation and hard x-ray emission. Schnurer *et al*. also obtained an x-ray emission decrease by reducing the pulse duration to a value smaller than 120 fs at constant laser energy and target position [23]. No explanation was given and it was suggested that it would need more experimental results. Therefore, in the regime of several tens of femtoseconds, further studies are still necessary to characterize the hard x-ray emission, especially the nature of the energy absorption mechanism.

In this paper, we study, theoretically, the possibility of x-ray source optimization from one and multilayer targets in both cold and hot compressed matters. We concentrate on Kα emission, since recently reported x-ray optic instrumentations make this particularly interesting. Firs we derive a general formula for the Kα x-ray yield from one layer laser-irradiated foils. Both forward and backward directions have been considered. Then this model is programmed by preparing a numerical code package. After that, this model is generalized for two and three layer targets and their results have been demonstrated. Finally, we generalized the so called model for fast electron propagation and Kα emission from the targets being heated and ionized by shock wave compression. These numerical results for multilayered cold and shocked targets are quite novel and have not performed in previous studies.

## II. ANALYTICAL MODEL OF Kα X-RAY GENERATION IN ONE-LAYER TARGETS

In order to understand our simulation results more clearly, an analytical model for plasma emission has been developed. When a metal target is irradiated by short laser pulses, hot electrons would be created on the surface of the target. These energetic electrons are directed into the slab along the target normal. This assumption is reasonable for non-relativistic electrons, regardless of the angle of incidence of the laser beam. In relativistic scheme, the penetration angle becomes energy dependent [24]. Recent experiments with short-pulse lasers indicate that distribution function for these hot electrons is assumed to be quasi Boltzmann [24],

$$f(E_0) = \frac{1}{T_h}\exp(-E_0/T_h) \quad (1)$$

Where $E_0$ is the initial energies of hot electrons at the surface and $T_h$ is their temperature in KeV.

The total number of hot electrons $N_h$ and their temperature $T_h$ at the surface of the target are related to the laser intensity $I$ as below [13]:

$$N_h(x=0) = \int_0^\infty N_h(x=0,E_0)dE_0 \approx 1.9\times10^{20} I^{-\frac{1}{2}} \quad (2)$$

$$kT_h \approx 130\sqrt{\frac{I}{10^{17}wcm^{-2}}} \ KeV \quad (3)$$

Figure 1 illustrated the Kα emission due to a distribution of $N_h$ hot electrons passing through a slab of thickness $d$. both forward and backward emission have been considered.

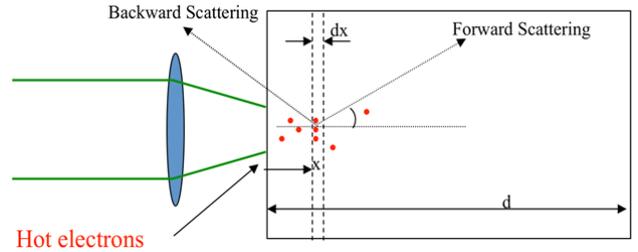

Fig. 1. Schematic of Kα emission from a one-dimensional slab target.

The number of Kα photons produced by hot electrons with energies between $[E, E+dE]$ within the interval $[x, x+dx]$ and emitted into a solid angle $\Omega$ at an angle $\theta$ relative to the electron penetration axis is calculated via [24]:

$$dN_{K\alpha} = N_h(x,E)n_A\sigma_k(E)\omega_K\frac{\Omega}{4\pi}\exp(-n_A\sigma_{ph}l)dEdx \quad (4)$$

Which $l$ is the oblique distance in which the Kα photons travel (at an angle θ) through the target.

$$0 \leq x \leq d \Rightarrow \begin{cases} l = \dfrac{x}{|\cos(\theta)|} & \text{Backward} \\ l = \dfrac{d-x}{\cos(\theta)} & \text{Forward} \end{cases} \quad (5)$$

where $N_h(x,E)$, $n_A$, $\sigma_k(E)$, $\omega_K$ and $\sigma_{ph}$ are, respectively, the number of hot electrons in depth $x$ with energy $E$, the atomic density of target material, K-shell ionization cross section, $K_\alpha$ fluorescence yield and photo-absorption cross section for the Kα photon.

$N_h(x,E)$ is proportion to number of electrons on the target surface as below:

$$N_h(x,E) = N_h(x=0) f(x, E_0 \to E) \quad (6)$$

It must be noted that the electrons with initial energies $E_0$ leave the target surface and reach the depth $x$ with final energies $E$ have a probability of $f(x, E_0 \to E)$. This function depends on the electron stopping power $dE/dx$. The stopping power related to various energies has been calculated from Ref. [25] which is based on NIST (National Institute for Standard & Technology), which electron energies from 1 $KeV$ to 10 $GeV$ are included.

It has been proved that initial energy of the electron $E_0$ and its energy $E$ at depth $x$, are connected by the below integral equation:

$$x = \int_E^{E_0} \dfrac{dE}{dE/dx} \quad (7)$$

By defining an angle-dependant photon mean free path $\lambda_{mfp}$ as:

$$\lambda_{mfp} = \dfrac{|\cos\theta|}{n_A \sigma_{ph}} \quad (8)$$

Total number of Kα photons which can reach the detector will be measured by integrating Eq.4 and 5 over x and E:

$$N_{K\alpha} = N_h n_A \omega_K \dfrac{\Omega_D}{4\pi} \int_0^\infty dE \sigma_K(E) \int_0^d dx\, f[E_0(E,x)] \quad (9)$$

$$\times \begin{cases} \exp\left(-\dfrac{x}{\lambda_{mfp}}\right) & \text{backward} \\ \exp\left(-\dfrac{d-x}{\lambda_{mfp}}\right) & \text{forward} \end{cases}$$

Withstanding the fact that the relationship among $x, E$ and $E_0$ is complicated, so solving Eq.9 become very difficult. Because the stopping power $\dfrac{dE}{dx}$ depend on $x$ and $E_0$, the two integrals in Eq.9 cannot be performed independently.

Nonetheless, by using some innovative techniques Eq.9 can be figure out.

Our simulation procedure is briefly consisted of three steps:

First, the target is divided into very thin layers of thickness $\Delta x (\Delta x \to 0)$. Then a three dimension matrix of $(x, E, E_0)$, which connect the energy $E$ of hot electrons at depth $x$ to their initial energy $E_0$ at front layer, is determined by use of electron stopping power $dE/dx$.

Second, the dimensionless quantities $Q_\pm$ regarding Eq.1 and Eq.9 are defined as below:

$$Q_+(E) = \int_0^d \dfrac{dx}{\lambda_{mfp}} \exp\left(-\dfrac{E_0(E,x)}{T_h}\right) \exp\left(-\dfrac{d-x}{\lambda_{mfp}}\right) \quad (10)$$

$$Q_-(E) = \int_0^d \dfrac{dx}{\lambda_{mfp}} \exp\left(-\dfrac{E_0(E,x)}{T_h}\right) \exp\left(-\dfrac{x}{\lambda_{mfp}}\right) \quad (11)$$

Integral equations 10 and 11 then have been calculated for a particular energy $E$ by applying the so called matrix $(x, E, E_0)$ in case of $E_0(x,E)$ in exponential terms.

Third, by considering Eq.10 and Eq.11, Eq.9 will be written as bellow:

$$N_{K\alpha} = N_h n_A \omega_K \dfrac{\Omega_D}{4\pi} \dfrac{\lambda_{mfp}}{T_h} \int_0^\infty dE \sigma_K(E) Q_\pm(E) \quad (12)$$

Performing step 2, this relation is then calculated by Simpson integration method.

We now proceed to analyze these equations with the aim of finding optimal conditions for emission as a function of target thickness, laser intensity and hot electron temperature. In the calculations that follow, the Kα emission cross sections, $\sigma_k$ from [26]; the fluorescence yields, $\omega_k$ from [27] and Kα photo-absorption cross section from [28]. It should be noted, however, that these parameters refer to cold materials. For compressed dense plasmas these have to be modified.

*A. Target thickness:*

In fig.2 the number of Kα photons per target thickness is shown. Temperature of hot electrons is assumed to be 40 KeV.

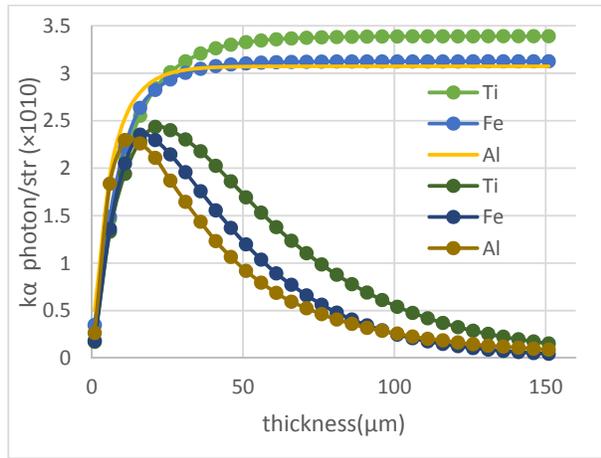

Fig. 2. Dependence of forward (solid lines) and backward (dashed lines) Kα emission on foil thickness for a three different targets ($T_h$=40 KeV).

It is observed that in forward scattering for all three types, there is a thickness which the Kα efficiency is maximized. For example, for Iron it is 15 micron. Also, in backward emission greater than a certain thickness, the Kα efficiency reaches a saturated state that doesn't increase.

*B. Laser intensity:*

Calculation for variation of laser intensity from $10^{14}$ to $10^{18}$ Wcm$^{-2}$ has been performed and simulation result illustrated below:

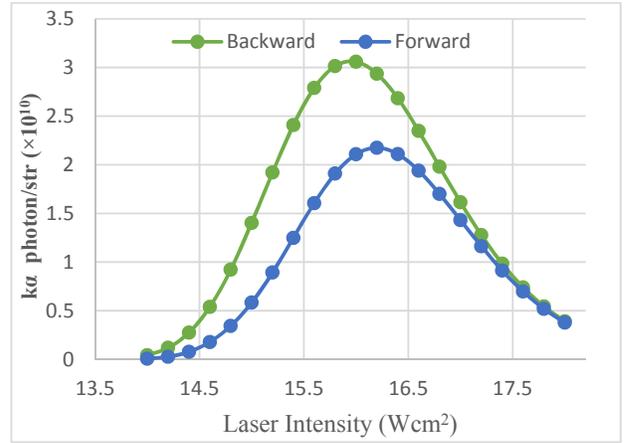

Fig. 3. Number of Kα x-ray emission as a function of laser intensity for both forward and backward scattering.

There is always this mentality that if laser intensity increase, the Kα x-ray emission rate will be enhanced. The results show that this notation is not necessarily true, but an optimum for laser intensity exists. It is because of photon reabsorption inside the target.

### III. MODEL OF Kα YIELD IN TWO LAYER TARGETS

Simulation of Kα x-ray for the metallic targets was mentioned in previous section. The generalization of single layer equation for two layer target according to Fig.4 is

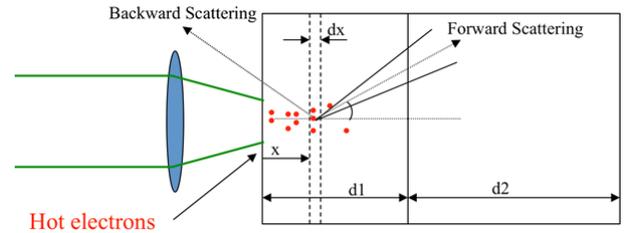

Fig. 4. Geometry of Kα emission from a two-dimensional slab target.

Following a similar approach we developed in section II, the number of Kα photons in the case of two-layered target can be formulated as follows,

$$N_{K\alpha} = [\int_{x=0}^{d1}\int_{E=0}^{\infty} N_h(x,E) n_1(Z_1)\sigma_1(Z_1,E)\omega_{K1}\frac{\Omega}{4\pi}$$
$$\exp(-n_1(Z_1)\sigma_{ph1}(Z_1,v)l)dEdx]$$
$$+[\int_{x=d1}^{d1+d2}\int_{E=0}^{\infty} N_h(x,E) n_2(Z_2)\sigma_2(Z_2,E)\omega_{K2}\frac{\Omega}{4\pi}$$
$$\exp(-n_2(Z_2)\sigma_{ph2}(Z_2,v)l)dEdx] \quad (13)$$

where

$$0 \le x \le d1 \Rightarrow \begin{cases} l = \dfrac{x}{|\cos(\theta)|} & \text{Backward} \\ l = \dfrac{d1+d2-x}{\cos(\theta)} & \text{Forward} \end{cases} \quad (14)$$

$$d1 \le x \le d1+d2 \Rightarrow \begin{cases} l = \dfrac{x}{|\cos(\theta)|} & \text{Backward} \\ l = \dfrac{d2-x}{\cos(\theta)} & \text{Forward} \end{cases} \quad (15)$$

Parameters with indices 1 are related to first layer and indices 2 for the second.

In forward scattering, the main purpose is the calculation of the first layer photons which are passed through the second layer. Hence, two points must be considered in simulation of Kα yield in two layer targets: generation of photons in the first layer and their penetration through the second layer.

In other words, the parameters in Eq.13, which are effective in production of Kα photons, are related to the first material. And the transmission parameters, related to both materials. The first layer parameters are used from the surface to thickness *d1* and the second layer parameters are used from *d1* to *d*.

On the other hand, in backward scattering, the numbers of Kα photons which are generated in the second layer and passed through the first layer should be considered.

*A. Simulation results of two layer targets:*

The plot of efficiency for a two-layer target which is made of titanium as first layer and iron as second layer is shown in fig.5.

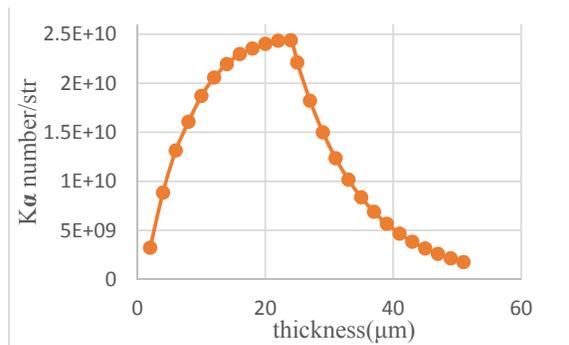

Fig. 5. Number of forward Kα photons from a 2-layer target Ti-Fe as a function of thickness. I=$10^{16}$ Wcm$^{-2}$ and T$_h$=40 KeV.

This figure is shown for the forward scattering according to the variation of the second layer thickness.

*B. Results for three layer target:*

In fig.6 the number of Kα forward emission of the first layer (Al) from a typical 3-layer target, composed of 12 micron Aluminum (first layer), 20 micron copper (second layer) and again 22 micron Aluminum, has been shown versus third layer thickness variation.

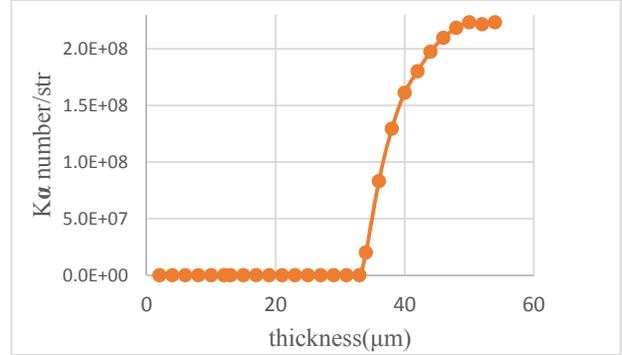

Fig. 6. Number of forward Kα photons emitted normally from a 3-layer Al-Cu-Al target versus third layer thickness in μm. I=$10^{16}$ Wcm$^{-2}$ and T$_h$=40 KeV.

Besides, the efficiency of Kα photons versus copper (2nd layer) thickness is given in fig.7.

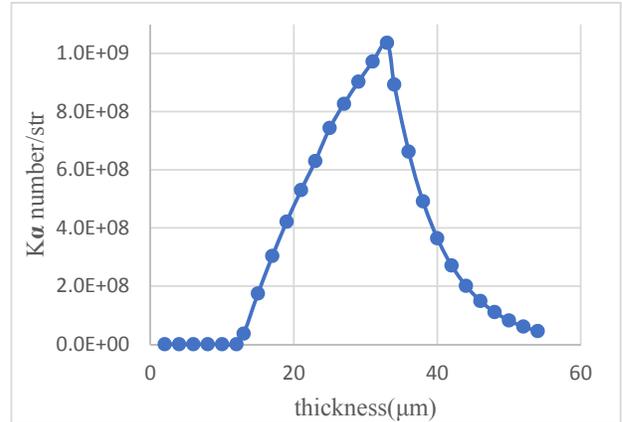

Fig. 7. Number of Cu-Kα photons emitted normally from 3-layer Al-Cu-Al target versus third layer thickness in μm. I=$10^{16}$ Wcm$^{-2}$ and T$_h$=40 KeV.

Clearly there is an overall optimal thickness— about 30 μm for emission in the forward direction (rearside of target).

## IV. SIMULATION OF SHOCKED COMPRESSED MATTERS

In inertial confinement fusion (ICF) research a new approach, the so called ''fast igniter'' was devised to achieve the goal of ignition [29,30]. Fast Ignition is a form of inertial fusion in which the ignition step and the compression step are separate processes. The invention of chirped pulse amplification2 of lasers spurred research in this area because these lasers can, in principle, supply energy to the fusion ignition region as fast as the convergence of stagnating flows can for the conventional ignition scheme [31-33].

Crucial to the success of this approach is to investigate the propagation of fast electrons and their energy deposition in the compressed pellet [34]. Furthermore, Kα x-ray emission, produced from the so called fast electrons is very important. In this section the model of Kα x-ray emission for cold multilayer targets was developed to cover targets heated and compressed by shock wave. To prove, our result was compared with the data of an experiment performed at the Rutherford Laboratory, by an Italian-French-English scientific team [35].

As illustrated in fig.8 a 3-layer target composed of 13/5 μm thick Chlorinated Plastic (PVCD) fluorescence layer which was sandwiched between $26\,\mu m$ of polyethylene ($CH_2$) on the side of the compression beams and 10–104 μm of polyethylene on the CPA side. In this experiment the propagation of the electrons through the target was studied using Kα spectroscopy by producing Kα emission from chlorine at 2622 eV. An ultra-fast pulsed laser with intensity of $10^{16}$ W cm$^{-2}$ is used to produce hot electrons on the front layer. In addition, by a laser source on the rear side, a shock wave in each layer is created. The shock wave cause to compress target and thus will increase the intensity of material. The experimental data showed a much larger fast electron penetration in the shocked material (about a 100% increase), which was due to the phase change produced in the target material by the shock wave compression [34,35].

The experiment showed a much increased penetration of fast electrons in the compressed plastic [35].

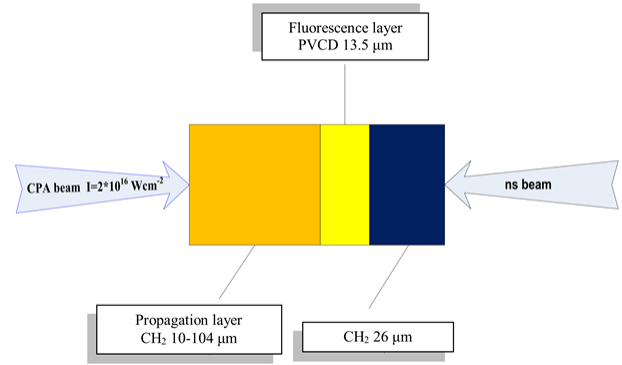

Fig. 8. In-principle scheme of the experimental setup and of the targets

Also a delay between the irradiation time CPA laser and the ns beam was chosen so that $\approx 8\,\mu m$ of uncompressed material was present when the CPA beam was fired. This ensured that the conditions of fast electron generation were identical in both the cold and shock-compressed cases.

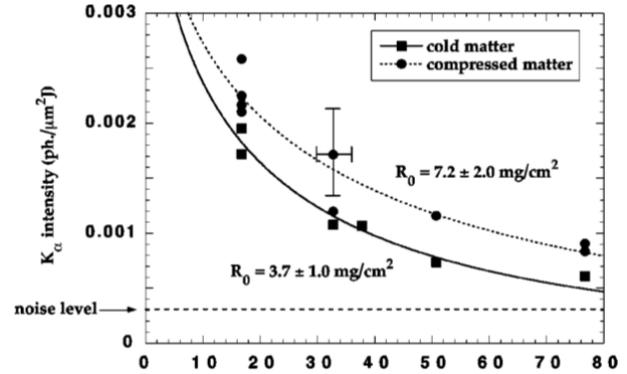

Fig. 9. Kα experimental yield and penetration depth for cold and compressed matter as a function of target thickness in μm [35]

The experimental data of Kα yield of Cl atom for both cold and compressed matter as a function of first layer target thickness are shown in fig.9.

### A. Results of compressed target; compare with experiment

In fig.10 number of Kα photons from Cl atom at 2.6 KeV in fluorescence layer (PVCD, 13.5 μm) was presented.

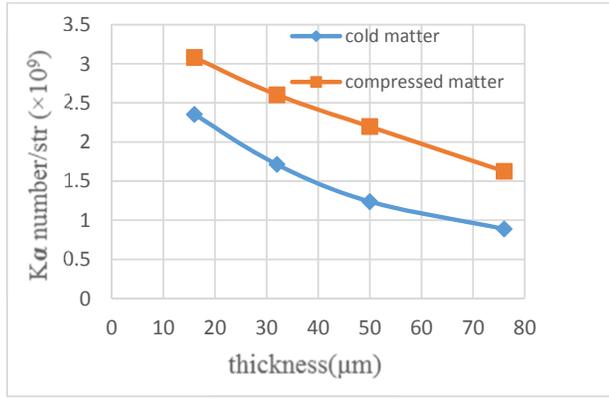

Fig. 10. Number of Kα photons for both cold and compressed matters as a function of first layer thickness in μm. I= $2\times10^{16}$ Wcm$^{-2}$ and $T_h$=40 KeV.

According to the results, using compressed target in case of cold target will greatly enhance the Kα efficiency that can be seen in both fig.9 and fig.10.

Comparing simulation results (fig.10) with the experimental results (fig.9), the accuracy of our model is confirmed. In the next 2 section number of Cl Kα photons versus 2nd and 3rd layer target thickness will be obtained.

*B. Second layer thickness*

In this section, the so called 3 layer target (CH$_2$-PVCD-CH$_2$) with 20 and 26 μm, respectively for first and third layer is considered and number of Cl Kα photons per second layer thickness is illustrated (fig.11)

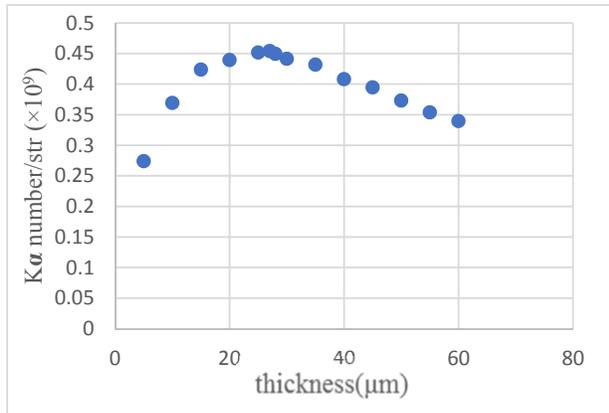

Fig. 11. Number of Cl-Kα photons as a function of second layer (PVCD) thickness in μm. I= $2\times10^{16}$ Wcm$^{-2}$ and $T_h$=40 KeV.

As seen in above figure, there is an optimum thickness for 2nd layer (27 micron in this case) which the number of Kα photons is maximized.

*C. Third layer thickness*

Now we stepped farther and studied Kα x-ray of Cl atom versus the third layer thickness variations, consuming the known 3-layer CH$_2$-PVCD-CH$_2$ target of 20 and 27 μm for the first and second layer respectively.

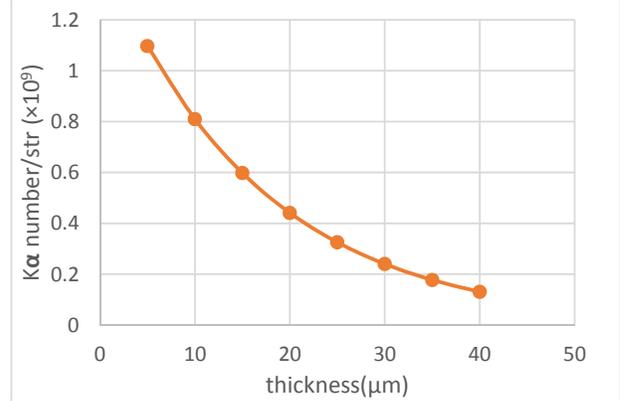

Fig. 12. Number of Kα photons as a function of third layer (CH$_2$) thickness in μm. I= $2\times10^{16}$ Wcm$^{-2}$ and $T_h$=40 KeV.

It shows that there is a descending trend for Kα with respect to third layer thickness.

## V. CONCLUSION

In summary, we have presented an analytical scaling model of Kα emission from multilayered targets which provide the optimum parameters for the incident laser and the target. A general model of Kα photon generation in femtosecond laser-irradiated solid material have been formulated which can be used for both cold and shocked compressed multilayer targets. Photon reabsorption is taken into account explicitly, allowing both forward and backscattered Kα emission to be simulated for arbitrary target thicknesses.

Moreover, we generalized this model and study the Kα fluorescence yield in shock compressed materials, and verified the accuracy of our model for both cold and hot compressed targets. Comparing the cold and the compressed cases, the number of detected Cl Kα photons for the compressed target is more than the cold one which is because of a much larger fast electron penetration depth and respectively electron stopping power in the shocked material (about a 100% increase), due to the phase change produced in the target material by the shock wave.

## REFERENCES

[1] P. Gibbon, *Short pulse laser interactions with matter*: World Scientific Publishing Company, 2004.
[2] G. A. Mourou, T. Tajima, and S. V. Bulanov, "Optics in the relativistic regime," *Reviews of Modern Physics,* vol. 78, pp. 309-371, Apr-Jun 2006.
[3] P. Mulser and D. Bauer, *High power laser-matter interaction* vol. 238: Springer Science & Business Media, 2010.
[4] E. G. Gamaly and A. V. Rode, "Physics of ultra-short laser interaction with matter: From phonon excitation to ultimate transformations," *Progress in Quantum Electronics,* vol. 37, pp. 215-323, Sep 2013.
[5] B. C. Stuart, M. D. Feit, S. Herman, A. M. Rubenchik, B. W. Shore, and M. D. Perry, "Nanosecond-to-femtosecond laser-induced breakdown in dielectrics," *Physical Review B,* vol. 53, pp. 1749-1761, 01/15/ 1996.
[6] E. Esarey, C. B. Schroeder, and W. P. Leemans, "Physics of laser-driven plasma-based electron accelerators," *Reviews of Modern Physics,* vol. 81, pp. 1229-1285, Jul-Sep 2009.
[7] B. Kadomtsev, *Reviews of plasma physics* vol. 13: Springer Science & Business Media, 2012.
[8] W. L. Kruer, *The physics of laser plasma interactions*, 1988.
[9] L. M. Chen, J. Zhang, Q. L. Dong, H. Teng, T. J. Liang, L. Z. Zhao, *et al.*, "Hot electron generation via vacuum heating process in femtosecond laser–solid interactions," *Physics of Plasmas,* vol. 8, pp. 2925-2929, 2001.
[10] A. Saemann, K. Eidmann, I. E. Golovkin, R. C. Mancini, E. Andersson, E. Forster, *et al.*, "Isochoric heating of solid aluminum by ultrashort laser pulses focused on a tamped target," *Physical Review Letters,* vol. 82, pp. 4843-4846, Jun 14 1999.
[11] D. Giulietti and L. A. Gizzi, "X-ray emission from laser-produced plasmas," *Rivista Del Nuovo Cimento,* vol. 21, pp. 1-93, 1998/10/01 1998.
[12] F. Ewald, H. Schwoerer, and R. Sauerbrey, "K-alpha-radiation from relativistic laser-produced plasmas," *Europhysics Letters,* vol. 60, pp. 710-716, Dec 2002.
[13] C. Reich, P. Gibbon, I. Uschmann, and E. Förster, "Yield Optimization and Time Structure of Femtosecond Laser K-alpha Sources," *Physical Review Letters,* vol. 84, pp. 4846-4849, 05/22/ 2000.
[14] T. Feurer, A. Morak, I. Uschmann, C. Ziener, H. Schwoerer, C. Reich, *et al.*, "Femtosecond silicon K-alpha pulses from laser-produced plasmas," *Physical Review E,* vol. 65, p. 016412, 12/19/ 2001.
[15] T. Pfeifer, C. Spielmann, and G. Gerber, "Femtosecond x-ray science," *Reports on Progress in Physics,* vol. 69, pp. 443-505, Feb 2006.
[16] A. Couairon and A. Mysyrowicz, "Femtosecond filamentation in transparent media," *Physics Reports-Review Section of Physics Letters,* vol. 441, pp. 47-189, Mar 2007.
[17] H. S. Park, D. M. Chambers, H. K. Chung, R. J. Clarke, R. Eagleton, E. Giraldez, *et al.*, "High-energy Kα radiography using high-intensity, short-pulse lasers," *Physics of Plasmas,* vol. 13, p. 056309, 2006.
[18] J. C. Kieffer, A. Krol, Z. Jiang, C. C. Chamberlain, E. Scalzetti, and Z. Ichalalene, "Future of laser-based X-ray sources for medical imaging," *Applied Physics B-Lasers and Optics,* vol. 74, pp. S75-S81, Jun 2002.
[19] J. Yu, Z. Jiang, J. C. Kieffer, and A. Krol, "Hard x-ray emission in high intensity femtosecond laser-target interaction," *Physics of Plasmas,* vol. 6, pp. 1318-1322, Apr 1999.
[20] B. Soom, H. Chen, Y. Fisher, and D. D. Meyerhofer, "Strong Kα emission in picosecond laser-plasma interactions," *Journal of Applied Physics,* vol. 74, pp. 5372-5377, 1993.
[21] D. C. Eder, G. Pretzler, E. Fill, K. Eidmann, and A. Saemann, "Spatial characteristics of K α radiation from weakly relativistic laser plasmas," *Applied Physics B: Lasers and Optics,* vol. 70, pp. 211-217, 2000/02/01 2000.
[22] A. Zhidkov, A. Sasaki, T. Utsumi, I. Fukumoto, T. Tajima, F. Saito, *et al.*, "Prepulse effects on the interaction of intense femtosecond laser pulses with high-\textit{Z} solids," *Physical Review E,* vol. 62, pp. 7232-7240, 11/01/ 2000.
[23] M. Schnurer, R. Nolte, A. Rousse, G. Grillon, G. Cheriaux, M. P. Kalachnikov, *et al.*, "Dosimetric measurements of electron and photon yields from solid targets irradiated with 30 fs pulses from a 14 TW laser," *Phys Rev E Stat Phys Plasmas Fluids Relat Interdiscip Topics,* vol. 61, pp. 4394-401, Apr 2000.
[24] D. Salzmann, C. Reich, I. Uschmann, E. Förster, and P. Gibbon, "Theory of K-alpha generation by femtosecond laser-produced hot electrons in thin foils," *Physical Review E,* vol. 65, p. 036402, 02/08/ 2002.
[25] M. J. Berger, J. Coursey, M. Zucker, and J. Chang, *Stopping-power and range tables for electrons, protons, and helium ions*: NIST Physics Laboratory, 1998.
[26] E. Casnati, A. Tartari, and C. Baraldi, "An empirical approach to K-shell ionisation cross section by electrons," *Journal of Physics B: Atomic and Molecular Physics,* vol. 15, p. 155, 1982.
[27] W. Bambynek, B. Crasemann, R. W. Fink, H. U. Freund, H. Mark, C. D. Swift, *et al.*, "X-Ray Fluorescence Yields, Auger, and Coster-Kronig Transition Probabilities," *Reviews of Modern Physics,* vol. 44, pp. 716-813, 10/01/ 1972.
[28] M. O. Krause, "Atomic radiative and radiationless yields for K and L shells," *Journal of Physical and Chemical Reference Data,* vol. 8, p. 307, 1979.
[29] E. Martinolli, M. Koenig, S. D. Baton, J. J. Santos, F. Amiranoff, D. Batani, et al., "Fast-electron transport and heating of solid targets in high-intensity laser interactions measured by K-alpha fluorescence," *Physical Review E,* vol. 73, p. 046402, 04/17/ 2006.
[30] J. Lindl, "Development of the indirect-drive approach to inertial confinement fusion and the target physics basis for ignition and gain," *Physics of Plasmas,* vol. 2, p. 3933, 1995.
[31] M. Tabak, D. S. Clark, S. P. Hatchett, M. H. Key, B. F. Lasinski, R. A. Snavely, *et al.*, "Review of progress in Fast Ignition," *Physics of Plasmas,* vol. 12, p. 057305, 2005.
[32] R. Kodama, P. A. Norreys, K. Mima, A. E. Dangor, R. G. Evans, H. Fujita, *et al.*, "Fast heating of ultrahigh-density plasma as a step towards laser fusion ignition," *Nature,* vol. 412, pp. 798-802, Aug 23 2001.
[33] M. Cerchez, R. Jung, J. Osterholz, T. Toncian, O. Willi, P. Mulser, *et al.*, "Absorption of ultrashort laser pulses in strongly overdense targets," *Phys Rev Lett,* vol. 100, p. 245001, Jun 20 2008.
[34] J. J. Santos, D. Batani, P. McKenna, S. D. Baton, F. Dorchies, A. Dubrouil, *et al.*, "Fast electron propagation in high density plasmas created by shock wave compression," *Plasma Physics and Controlled Fusion,* vol. 51, p. 014005, Jan 2009.
[35] D. Batani, J. R. Davies, A. Bernardinello, F. Pisani, M. Koenig, T. A. Hall, *et al.*, "Explanations for the observed increase in fast electron penetration in laser shock compressed materials," *Physical Review E,* vol. 61, pp. 5725-5733, 05/01/ 2000.